%% file: main_IPDPS_2022.tex
\documentclass[conference]{IEEEtran}
\IEEEoverridecommandlockouts
\usepackage{cite}
\usepackage{amsmath,amssymb,amsfonts}
\usepackage{graphicx}
\usepackage{textcomp}
\usepackage{xcolor}

\usepackage{comment}
\usepackage{cleveref}
\usepackage[linesnumbered,ruled,vlined]{algorithm2e}
\usepackage{algpseudocode} 
\usepackage{framed}
\usepackage{mwe}
\usepackage{pgffor}
\usepackage{xcolor}
\SetCommentSty{mycommfont}
\SetKwInput{KwInput}{Input}                
\SetKwInput{KwOutput}{Output}              
\SetKwInput{KwInitialize}{Initialize}              

\def\BibTeX{{\rm B\kern-.05em{\sc i\kern-.025em b}\kern-.08em
    T\kern-.1667em\lower.7ex\hbox{E}\kern-.125emX}}
\begin{document}

\title{Job Scheduling in Datacenters using Constraint Controlled RL}

\author{\IEEEauthorblockN{Vanamala Venkataswamy}
\IEEEauthorblockA{\textit{Department of Computer Science} \\
\textit{University of Virginia}\\
Charlottesville, USA \\
vv3xu@virginia.edu}
}

\maketitle

\input{0-abstract}
\input{1-introduction}
\input{2-background}
\input{3-approach}  
\input{4-design}

\input{5-evaluation}

\input{7-related-work} 
\input{6-discussion}

\input{8-conclusion}

\bibliography{main_IPDPS_2022.bbl}

\end{document}

%% file: 0-abstract.tex
\begin{abstract}
This paper studies a model for online job scheduling in green datacenters. In green datacenters, resource availability depends on the power supply from the renewables. Intermittent power supply from renewables leads to intermittent resource availability, inducing job delays (and associated costs). Green datacenter operators must intelligently manage their workloads and available power supply to extract maximum benefits. The scheduler's objective is to schedule jobs on a set of resources to maximize the total value (revenue) while minimizing the overall job delay. A trade-off exists between achieving high job value on the one hand and low expected delays on the other. Hence, the aims of achieving high rewards and low costs are in opposition. In addition, datacenter operators often prioritize multiple objectives, including high system utilization and job completion. To accomplish the opposing goals of maximizing total job value and minimizing job delays, we apply the Proportional-Integral-Derivative (PID) Lagrangian methods in Deep Reinforcement Learning to job scheduling problem in the green datacenter environment. Lagrangian methods are widely used algorithms for constrained optimization problems. We adopt a controls perspective to learn the Lagrange multiplier with proportional, integral, and derivative control, achieving favorable learning dynamics. Feedback control defines cost terms for the learning agent, monitors the cost limits during training, and continuously adjusts the learning parameters to achieve stable performance. Our experiments demonstrate improved performance compared to scheduling policies without the PID Lagrangian methods. Experimental results illustrate the effectiveness of the \textbf{Co}nstraint \textbf{Co}ntrolled \textbf{R}einforcement \textbf{L}earning (CoCoRL) scheduler that simultaneously satisfies multiple objectives.

\end{abstract}


%% file: 1-introduction.tex

\section{Introduction and Motivation}

The exponential growth in demand for digital services drives massive datacenter energy consumption and negative environmental impacts. Promoting sustainable solutions to pressing energy and digital infrastructure challenges is crucial. Several hyperscale cloud providers have announced plans to power their datacenters using renewable energy. However, integrating renewables to power the datacenters is challenging because the power generation is intermittent, necessitating approaches to tackle power supply variability. Hand engineering domain-specific heuristics-based schedulers to meet specific objective functions in such complex dynamic green datacenter environments is time-consuming, expensive, and requires extensive tuning by domain experts. The green datacenters need smart systems and system software to employ multiple renewable energy sources (wind and solar) by intelligently adapting computing to renewable energy generation. 

Reinforcement Learning has solved sequential decision tasks of impressive difficulty by maximizing reward functions through trial and error. Recent examples using deep learning range from robotic locomotion~\cite{robotics_2015}~\cite{robotics_2016}, sophisticated video games~\cite{games_2018}~\cite{OpenAI}, congestion control~\cite{congestion_control}, and cluster job scheduling~\cite{RLScheduler-SC20}~\cite{GreenScheduler-MLCS20}~\cite{RLScheduler-JSSPP2022}. While errors during training in these domains come without a cost, limiting the rates of hazardous outcomes in some learning scenarios is crucial. One example is wear and tear on a robot's components or surroundings. It is possible to impose such limits directly by prescribing constraints in the action or state space; hazard-avoiding behavior must be learned. 

Models trained using Reinforcement Learning (RL) are a powerful replacement for hand-written heuristics. Existing RL cluster schedulers presented in  ~\cite{google-spotlight}~\cite{Metis}~\cite{DeepEE} solely focus on optimizing a single reward. While these implementations demonstrate good performance for a specific metric, they do not consider the cost associated with hazardous actions. In practice, these RL policies face skepticism about their robustness when encountering unusual situations. For example, an RL model may "misbehave" on an unanticipated workload change or intermittent power supply making bad scheduling decisions that lead to applications' service level objectives (SLOs) violations. 

\begin{figure}[h]
\centering
\includegraphics[width=\linewidth]{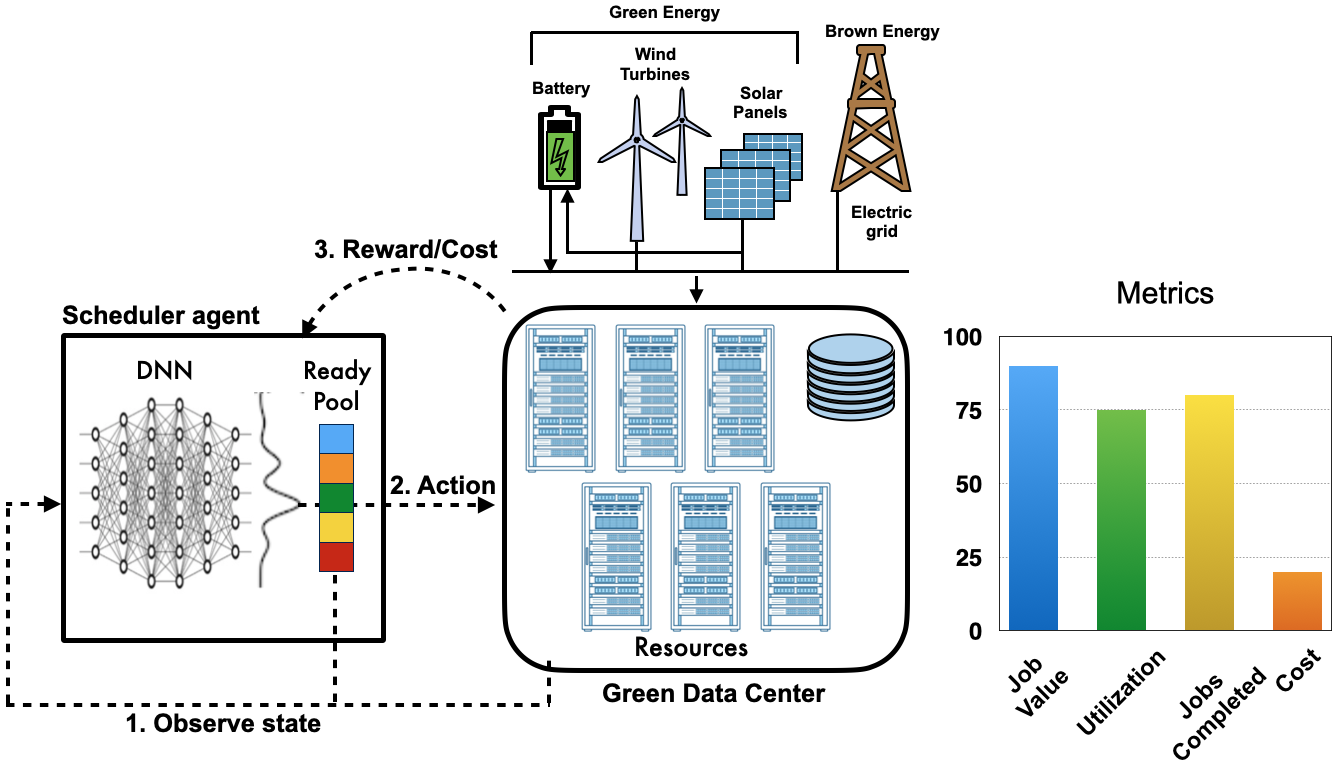}
\caption{DRL scheduler interacting with Green datacenter (powered by renewable and brown energy sources).}
\label{fig:gdc-model}
\end{figure}


Additionally, focusing only on generating "good" or "balanced" schedules is not enough. The agent must not only satisfy the constraints but also optimize for other metrics (Fig.\ref{fig:gdc-model}). The two most common objectives that datacenter operators prioritize are order-based and resource-based. For example, satisfying due dates (reducing job delays) is an order-based objective, while efficient resource utilization is a resource-based objective. This can be achieved by appropriately co-designing the reward and cost functions to satisfy multiple objectives simultaneously.

For this purpose, we adapt the well-known framework of the constrained Markov decision process (CMDP)~\cite{altman-98}, which limits the accumulation of a "cost" signal which is analogous to the reward. The desired policy is one that maximizes the expected return while satisfying the cost constraint. 

Furthermore, any effective scheduling system must be able to adapt to unforeseen events. In a green datacenter environment, the resource availability is based on the power supply from renewables. Even with the best weather predictions, there can be unforeseen circumstances when the power supply drops drastically. The scheduler should handle such situations gracefully by maintaining high rewards and costs within limits.

One solution is to embed all conflicting requirements in a constrained RL problem and use a primal-dual algorithm that automatically chooses the agent's parameters. The main advantage of this approach is that constraints ensure satisfying behavior without manually selecting the penalty coefficients. Instead of applying constraints in the action or state-space directly, hazard-avoiding behavior is learned. We utilize the well-known framework of the Constrained Markov Decision Process (CMDP)~\cite{altman-98}, limiting the accumulation of the cost signal, which is similar to the reward signal. Lagrangian methods are a classic technique for solving constrained optimization problems. The desired scheduling policy is one that maximizes the usual return while satisfying the cost constraint. 


Our contributions are outlined as follows. 
\begin{itemize}
\item We present a Constraint-controlled RL (CoCoRL) scheduler that uses a primal-dual algorithm to automatically learn conflicting reward and cost functions. 
\item We demonstrate that a Constraint-controlled RL scheduler learns policies that not only satisfy the constraints but also optimize for other objectives such as resource utilization and job completion. 
\item Our results illustrate that our CoCoRL scheduler is robust and efficiently adapts to real HPC workloads while using real power supply data (solar and wind) from an existing Green datacenter.
\item We illustrate the importance of accurately tuning hyperparameters to satisfy various optimization goals set by datacenter operators.
\end{itemize}

The rest of the paper is organized as follows. In \S\ref{background}, we present background in \S \ref{background} and our approach in \S \ref{approach}. In \S\ref{crl_scheduler_design}, we introduce DRL framework and the Constrained-controlled RL scheduler design. We evaluate the scheduler and present the results in \S\ref{evaluation} followed by related work in \S\ref{related_work}. We present future work in \S\ref{discussion} and the conclusions in \S \ref{conclusion}.

%% file: 2-background.tex

\section{Background} \label{background}

In many situations in the optimization of dynamic systems, a single utility for the optimizer may not be sufficient to describe all the objectives involved in sequential decision-making. A special situation where one controller has multiple objectives. Instead of introducing a single utility to maximize (or a cost minimize) that may be some function (e.g., some weighted sum) of the multiple objectives, a natural approach for handling such cases is optimizing one objective with constraints on others. In particular, this allows us to understand the trade-off between the various objectives. 

\begin{figure}[h]
\centering
\includegraphics[scale=0.55]{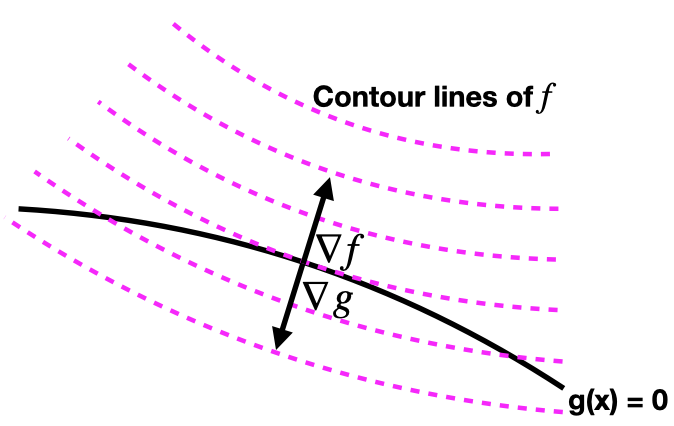}
\caption{The constrained minimum, $\nabla f = \text{-} \lambda \nabla g$.}
\label{fig:f-g-constraints}
\end{figure}

One solution is to embed all conflicting requirements in a constrained RL problem and use a primal-dual algorithm that automatically chooses the agent's parameters (Fig.\ref{fig:f-g-constraints}). This approach's main advantage is that constraints ensure satisfying behavior without manually selecting the penalty coefficients. Instead of applying constraints in the action or state-space directly, hazard-avoiding behavior is learned. We utilize the well-known framework of the Constrained Markov Decision Process (CMDP)~\cite{altman-98}, limiting the accumulation of the cost signal, which is similar to the reward signal. The optimal scheduling policy is one that maximizes the usual return while satisfying the cost constraint. 

\subsection{Lagrangian Methods for Constrained Optimization} 

Lagrangian methods are a classic family of approaches to solving constrained optimization problems. For example, the equality-constrained problem over the real vector $\mathrm{x}$:

\begin{equation} \label{eq:lagrangian-eq}
\min_{\mathrm{x}} f(\mathrm{x}) \ \ \ \text{s.t.}\ g(f) = 0 
\end{equation}

is transformed into an unconstrained one by introduction of a dual variable–the Lagrange multiplier, $\lambda$–to form the Lagrangian: $\mathcal{L}(\mathrm{x}, \lambda) = f(\mathrm{x}) + \lambda g(\mathrm{x})$, which is used to find the solution as:

\begin{equation} \label{eq:lagrangian-dual}
(\mathrm{x}^{\ast}, \lambda^{\ast}) = \arg\max_{\mathrm{x}}\ \min_{\lambda}\ \mathcal{L}(\mathrm{x}, \lambda)
\end{equation}

Gradient-based algorithms iteratively update the primal and dual variables where $\lambda$ acts as a learned penalty coefficient in the objective, leading to a constraint-satisfying solution~\cite{bertsekas-2014}. 

\subsection{Dynamical Systems and Feedback Control} 
Dynamical systems are processes that are subject to external control. A generic formulation for discrete-time systems with feedback control is:
\begin{equation} \label{eq:dynamical-system}
\begin{split}
\mathrm{x}_{k+1} = F(\mathrm{x}_k, u_k) \\
\mathrm{y}_k = Z(\mathrm{x}_k) \\
u_k = h(\mathrm{y}_0, \mathrm{y}_1, \ldots, \mathrm{y}_k) \\
\end{split}
\end{equation}
Where $\mathrm{x}$ is the state vector, $F$ is the dynamics function, $u$ applied control, $\mathrm{y}$ is the measurement outputs, and the subscript denotes the time step. The feedback rule, $h$, can access past and present measurements. The optimal control problem is to design a control rule, $h$, that results in a sequence $\mathrm{y}_{0:T} = \{\mathrm{y}_0, \mathrm{y}_1, \ldots, \mathrm{y}_T\}$ (or states $\mathrm{x}_{0:T}$) that scores well for some cost function $C$. For instance, reaching a goal condition, $C = \vert \mathrm{y}_T\ \mathrm{-}\ \bar{\mathrm{y}} \vert$, or closely following a desired trajectory, $\bar{\mathrm{y}}_{0:T}$.

\begin{figure}[h]
\centering
\includegraphics[width=\linewidth]{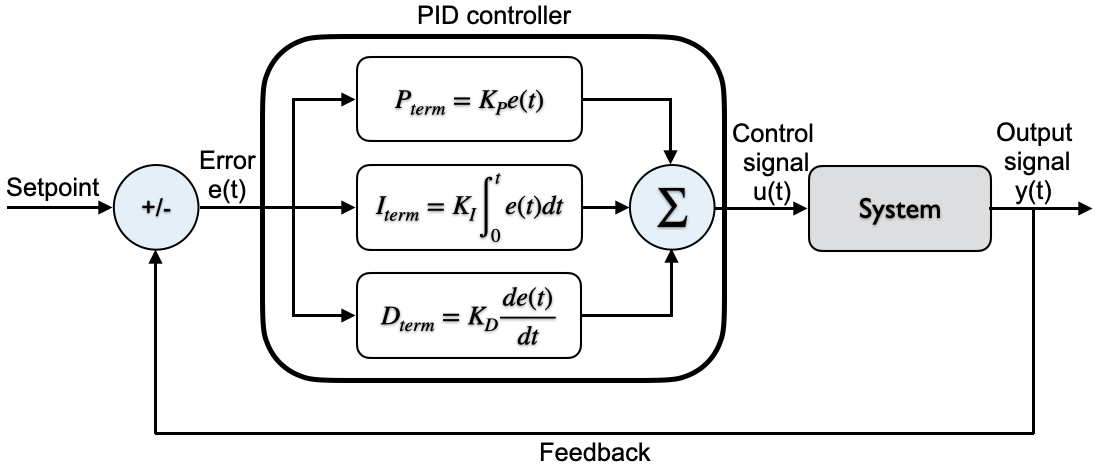}
\caption{Typical PID feedback loop to control the system.}
\label{fig:pid_controller}
\end{figure}

A typical feedback control system, illustrated in Fig.\ref{fig:pid_controller}, is composed of a controller, a system to be controlled, actuators, and sensors. The setpoint represents the exact value of the controlled variable. The error is the difference between the setpoint and the current value of the controlled variable, i.e., $e(t) = setpoint - current\ value$ of the controlled variable. The manipulated variable is the quantity that the controller varies to influence the value of the controlled variable. The feedback loop of the system is as follows: 1) The system, at regular intervals, monitors and compares the controlled variable to the setpoint to determine the error, 2) The controller computes the required control signal based on the error; and 3) The actuators change the value of the manipulated variable to control the system. 

\subsection{Constrained Reinforcement Learning} \label{cmdp}
The Constrained Markov Decision Processes (CMDP)~\cite{altman-98} extend MDPs~\cite{sutton-barto-1998} to include constraints into RL. A CMDP is the extended tuple $(S, A, R, T, \mu, C_0, C_1, \ldots, d_0, d_1, \ldots)$. The cost functions $C_i : S \times A \times S \rightarrow \mathbb{R}$ defined with same form as the reward functions, and $d_i : \mathbb{R}$ representing cost limits. For this work, we will consider a single, all-encompassing cost.

In RL, the expected sum of discounted rewards computed over $\tau = (s_0, a_0, r_1, s_1, a_1, r_2, \ldots)$ trajectories, using the policy $\pi(a \vert s)$ is a common performance objective: $J(\pi) = E_{\tau \sim \pi} [ \sum_{t=0}^{\infty} \gamma^t R(s_t, a_t, s_{t+1})]$. The corresponding value function for the cost is defined as: $J_C(\pi) = E_{\tau \sim \pi} [ \sum_{t=0}^{\infty} \gamma^t C(s_t, a_t, s_{t+1})]$. Then the constrained RL problem is solved for the best possible policy: 
\begin{equation} \label{eq:value-fun-cost}
\pi^{\ast} = \arg\max_{\pi} J(\pi)\ \ \ \text{s.t.}\ J_C(\pi) \leq d 
\end{equation}

DRL uses a Deep Neural Network (DNN) for the policy, $\pi_{\theta} = \pi(\cdot \vert s; \theta)$ with $\theta$ as parameter vector. The policy gradient algorithms improve the policy over time by gathering experience in the task of estimating the reward objective gradient, $\nabla_{\theta}J(\pi_{\theta})$ iteratively. Therefore our constrained optimization problem is expressed as maximizing score at some iterate, $\pi_k$, ideally obeying constraints at each iteration:
\begin{equation} \label{eq:min-max}
\begin{split}
\max_{\pi} J(\pi_k)\ \ \text{s.t.}\ J_C(\pi_m) \leq d \\
\text{where } m \in \{0, 1, \ldots, k\}
\end{split}
\end{equation}

We cast constrained RL as a dynamical system with the Lagrange multiplier as a control input, to which we apply PID control in the learning algorithm. The PID multiplier method proposed in~\cite{adam-2020} is a recent result where a PID update rule is considered for a learned Lagrange multiplier.

%% file: 3-approach.tex

\section {Approach} \label{approach}
This section presents the mapping of RL as a dynamical system and the PID Lagrangian method for constrained-controlled RL agent. In section \ref{crl_scheduler_design}, we present scheduler design as an instance of this mapping and the RL training algorithm.

\subsection{Mapping RL as a Dynamical System}
Similar to the system of equations in eq.\ref{eq:dynamical-system}, the first-order dynamical system in constrained RL ~\cite{adam-2020} form is defined as:
\begin{equation} \label {eq:dynamic_system}
\begin{split} 
\theta_{k+1} = F(\theta_k, \lambda_k) \\
y_k = J_C(\pi_{\theta_k}) \\
\lambda_k = h(y_0, y_1, \ldots, y_k, d) \\
\end{split}
\end{equation}
Here $F$ is a nonlinear function corresponding to the policy update on the RL agent's parameter vector, $\theta$. The penalty or cost-objective, $y$, is the measured output of the system. This measured output ($y$) is fed to the feedback control rule, $h$, along with the cost limit, $d$. With this starting point, the learning algorithm given by $F$, and the penalty coefficient update rule, $h$, can be tailored to solve the constrained optimization in (eq.\ref{eq:min-max}).

The policy gradients for reward and cost of the first-order Lagrangian method, $\nabla_{\theta} \mathcal{L}(\theta, \lambda) = \nabla_{\theta}J(\pi_{\theta}) \mathrm{-} \lambda \nabla_{\theta}J_C(\pi_{\theta})$, is organized in the form of eq.\ref{eq:dynamic_system} as:

\begin{equation}
F(\theta_k, \lambda_k) = f(\theta_k) + g(\theta_k)\lambda_k 
\end{equation}
\begin{equation}
f(\theta_k) = \theta_k + \eta \nabla \theta J(\pi_{\theta_k} ) 
\end{equation}
\begin{equation}
g(\theta_k) = \mathrm{-}\eta \nabla \theta J_C(\pi_{\theta_k} ) 
\end{equation}
where $\eta$ is the Stochastic Gradient Descent (SGD) learning rate. The controller's role is to push inequality constraint violations $(J_C \mathrm{-} d)_+$ to zero. 

\subsection{Lagrangian update with PID controller}
A constrained optimization problem is a problem of the form maximize (or minimize) the function $F(x,y)$ subject to the condition $g(x,\ y) = 0$. The Constrained MDPs have two criteria; 1) the usual reward and 2) the cost as a second value function. The reward must be optimized while the cost must remain below some specified threshold.

The PID update rule,~\cite{adam-2020}, is shown in Algorithm.\ref{algo:pid_update}. The proportional term fastens the response to constraint violations, the integral term eliminates steady-state violations at convergence, and the derivative control acts in anticipation of violations. It prevents cost overshoot and limits the rate of cost increases within the feasible region. 

\begin{algorithm}[h]
	\DontPrintSemicolon
	\KwInitialize {
	    $J_{C,prev} \leftarrow 0$ \Comment{Previous Cost} \\
        $I \leftarrow 0$ \Comment{Integral} \\
        $K_p, K_i, K_d > 0$ \Comment{Tuning parameters}
    }
    \For {i,\ldots,i+k} {
        Collect current cost, $J_C$ \\
        $e(t) = J_C \mathrm{-} d$  \\
        \tcp{proportional term}
        $P = K_p e(t)$ \\
        \tcp{integral term}
        $I = (I + e(t))_+$ \\
        \tcp{differential term}
        $D = K_d (J_C - J_{C,prev})_+$  \\
        \tcp{Control output}
        $\lambda = (K_p P + K_i I + K_d D)_+$ \\
        $J_{C,prev} = J_C$ \\
        return $\lambda$ 
    }
\caption{Update Lagrange Multiplier using PID control parameters.} 
\label{algo:pid_update}
\end{algorithm}

%% file: 4-design.tex

\section {Constrained-controlled RL Scheduler Design} \label{crl_scheduler_design}
This section presents the green datacenter overview, workload, and resource model. Next, we will discuss the Constraint-controlled RL scheduler and training algorithm overview.

\subsection{Green Datacenter Overview}
The green datacenter is a datacenter co-located at or near renewable energy sources. Various renewable sources can power the datacenter with the provision to store (battery) excess energy from renewables. Additionally, the datacenter is connected to the electric grid to support critical infrastructure when energy from renewables and batteries cannot sustain the load. We aim to design a green datacenter environment that can be controlled by heuristic and RL-based scheduling policies. In order to train the RL scheduler, we convert the datacenter scheduling problem into a CMDP(\S\ref{cmdp}, ~\cite{altman-98}) with a state space $\mathbf{S}$ describing the current status of the cluster resources, an action space $\mathbf{A}$ of new jobs, and a reward function $\mathbf{R}$ to be optimized. The operation of the datacenter - including receiving new jobs and placing scheduled jobs on available resources - becomes the MDP transition function, $\mathbf{T}$. Fig. \ref{fig:gdc-model} provides an overview of a DRL scheduler agent interacting with the green datacenter environment. 

\subsection{State Space, Reward and Cost Functions} \label{state_space}
The state space, $\mathcal{S}$, includes information about jobs, resources, and resource availability (based on power generation predictions). 

\subsubsection{Resources} \label{resources}
Our green datacenter simulator models a pool of servers (CPUs and GPUs), enabling the scheduler to make granular per-resource scheduling decisions. The available resources are allocated contiguously to the jobs. The power availability feedback is not directly provided to the scheduler agent. Instead, the resource pool expands and contracts based on the power available at the datacenter at any given time. Power availability decides when and how many resources are turned on or off. Therefore, power prediction data is an integral part of the state space, i.e., as power availability changes, the corresponding resource availability is reflected in the state information supplied to the scheduler agent. 

\subsubsection{Jobs} \label {jobs_qos}
In our system, jobs can be in one of three locations: 1) wait\_pool, 2) ready\_pool, or 3) scheduled on the resources. The wait\_pool is where jobs first arrive. The jobs from wait\_pool are moved to the ready\_pool, where they can then be scheduled on the resources. Jobs have meta-data, including the job's id, value, and resource requirements. 

The jobs are processed over fixed T timesteps. The time-horizon shifts after processing jobs during that timestep, with the job metadata vectors updated and the resource image advancing by one row. As the time-horizon shifts, the power supply from renewables (and batteries) dictates the availability of resources. The scheduler agent continuously observes the state of jobs, resources, and resource/power availability to make scheduling decisions. 

\textit{QoS of a job}: Users may have different utility functions, i.e., users are willing to pay different amounts for different jobs based on their importance. The user picks the required QoS for that job based on the user's willingness to pay for the job (e.g., spot instances\cite{spot_amazon}). The QoS value is specified as a percentage of the time the user wants his job to run. The qos\_violation\_time, $\ (expected\_finish\_time \div Qos\ Value$), specifies the upper bound by which the job must finish executing. If a job remains in the system past qos\_violation\_time, it incurs negative rewards every time after that. The higher the QoS value, the closer the job's completion time to the expected\_finish\_time. Expressing QoS value in percentages gives an upper bound of when a user can expect his job to finish. The idea is similar to Least Attained Service (LAS)\cite{gavel} in that if preempted, a job that has received more service is suspended and later restarted.


\subsubsection{Reward and Cost Functions} \label{rewards}
The DRL scheduler's objective is realized with rewards that the agent receives. Rewards, which are scalars given by the reward function $\mathbf{R}(s_t, a_t, s_{t+1})$, are a combination of the positive reward or associated cost for the action in a given state. Some actions collect positive rewards, while other actions accrue negative costs. For instance, if a job, $j$, is running on a resource, it collects a positive reward proportional to the job's value. A job's value, $j._{value}$, is calculated based on the type of resources requested, duration, and QoS value. If a job is delayed and QoS is violated, it collects a negative reward. Negative reward indirectly encourages fairness, ensuring low QoS value jobs are not delayed or starved. Other costs and rewards can be incorporated into the reward function. Our DRL scheduler's objective is to maximize the total job value from finished jobs, $\vert J_{finished} \vert$ expressed as,
\begin{equation} \label{eq:job_value}
Total\ Job\ Value = \sum_{i=1}^{\vert J_{finished}\vert} j_i.value 
\end{equation}
Rewards have a small and dense component encouraging movement toward the goal (completing jobs) at every step. A direct calculation of value is the price the user is willing to pay to run a job. Total Job Value is both an application-centric and resource-centric metric; the emphasis is on processing as many user jobs as possible, which may increase resource utilization. By processing as many jobs as possible, we essentially maximize the total value we gain from running those jobs. Even a small improvement in total job value can generate millions of dollars in savings for the service providers. 
\begin{equation} \label{eq:cost}
Cost = \sum_{i=1}^{n} \mathbf{1}_{j_i \in J}
\end{equation}

The environment provides a separate cost signal for each delayed job at every timestep. These costs are separate from the task-based reward signal. We can customize the costs to be an aggregate cost signal reflecting more than one constraint, e.g., job delays and QoS violations. By default, cost functions are simple indicators of whether a job is delayed or not. The cost signal, waiting or delayed jobs, is expressed in eq.\ref{eq:cost}.

\subsection{Constraint-controlled RL Scheduler} \label{crl_scheduler}
\begin{figure}[htbp]
\centering
\includegraphics[width=\linewidth]{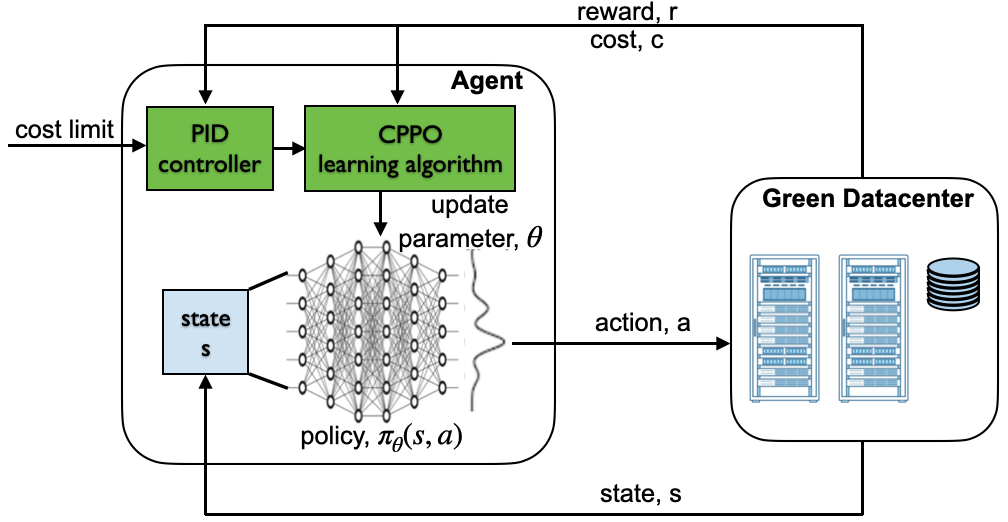}
\caption{DRL scheduler with PID controller (highlighted in green) interacting with the green datacenter. A Constraint-controlled policy ensures the system actions generated by the policy neural network do not violate constraints during training and deployment} 
\label{fig:drl_pid_scheduler}
\end{figure}

Fig.\ref{fig:drl_pid_scheduler} illustrates the CoCoRL scheduler overview.  We applied the Constrained-controlled Proximal Policy Optimization (CPPO)~\cite{adam-2020} policy gradient method, a constraint-controlled variant of PPO~\cite{ppo-2017}. The training (Algorithm.\ref{algo:cppo-training}) follows the typical minibatch-RL scheme. The agent senses the current state, takes action, and records the reward information for a fixed number of time steps in each episode. The trajectory taken during each episode is recorded as, $\tau = <s_0, a_o, r_1, s_1, a_1, r_2, \ldots>$. Rewards are computed from recorded values for each time step $t$ of every episode. The sampled estimates of the cost criterion, $\hat{J}_C$, are fed back to control the Lagrange multiplier (refer to Algorithm ~\ref{algo:pid_update}). As the agent learns for rewards, the upward pressure on costs from reward learning can change, requiring a dynamic response. The costs are fed to the PID controller to control the Lagrange multiplier, $\lambda$. The $\theta-$learning loop in the CCPO method updates the parameters, $\theta$, accordingly.

\begin{algorithm}[h]
    \SetAlgoLined
    \DontPrintSemicolon
    \KwInput{Batch Size $B$, Learning Rate $\alpha$, Discount $\gamma$, Cost limit \\
    Datacenter Simulator Env with Dynamics $\mathbf{T} : \mathcal{S} \times \mathcal{A} \rightarrow \mathcal{S}$ \\ Reward Function $\mathbf{R} : \mathcal{S} \times \mathcal{A} \times \mathcal{S} \rightarrow \mathbb{R}$ \\ Cost Function $\mathbf{C} : \mathcal{S} \times \mathcal{A} \times \mathcal{S} \rightarrow \mathbb{C}$ }  
    \KwInitialize{Actor Net $\pi_{\theta}$, Critic Nets $V_{\phi}$, $V_C,\psi$,  \\
    Control rule, $J_C \leftarrow [\ ]$ //\small\texttt{cost history} } 
    \For {training step $t \in \{0,\dots,T\}$} {
    {
	    Sample the environment: mini-batch \\
	    {
	        \hskip1.0em \tcp{sample action}
		    \hskip1.0em $a \sim \pi(\cdot \vert s; \theta)$ \\
		    \hskip1.0em \tcp{transition to new state}
		    \hskip1.0em $s\;' \sim T(s, a)$ \\
		    \hskip1.0em \tcp{collect reward}
		    \hskip1.0em $r \sim R(s, a, s\;')$ \\
		    \hskip1.0em \tcp{collect cost}
		    \hskip1.0em $c \sim C(s, a, s\;')$ \\
		}
	    Apply feedback control: \\
		    \hskip1.0em Store sample estimate $\hat{J}_C$ into $J_C$ \\
		    \hskip1.0em $\lambda \leftarrow h(J_C, d), \lambda \geq 0$ //\small\texttt{see algo~\ref{algo:pid_update}} \\ 
	    Update $\pi$ using Lagrangian objective \\
	        \hskip1.0em Update critics, $V\phi(s), V_C,\psi(s)$ \\
	        \hskip1.0em \tcp{reward and cost policy gradients}
	        \hskip1.0em $\nabla_{\theta} L = \nabla_{\theta} \hat{J}(\pi_{\theta}) \mathrm{-} \lambda
	    \nabla_{\theta} \hat{J}_C(\pi_{\theta})$ \\
	}
    }
    \KwOutput{Trained Scheduling Policy $\pi_{\theta}$}
\caption{CoCoRL training with Constrained PPO policy gradient method.} 
\label{algo:cppo-training}
\end{algorithm}

%% file: 5-evaluation.tex
\section {Evaluation} \label{evaluation}

We investigated the performance of the CoCoRL scheduling algorithm, the CPPO~\cite{adam-2020} policy gradient method described in Algorithm ~\ref{algo:cppo-training}, on a small green datacenter setting with ten resources. The policy network is a two-layer Multilayer Perceptron (MLP) and a final Long Short-Term Memory (LSTM)~\cite{lstm}. The reward function (the primary objective) maximizes the total job value, while the cost function (the secondary objective) limits the overall job delay. We show the effectiveness of PID control in maximizing the total value while simultaneously reducing constraint violations. The baseline heuristic scheduling policies, by design, do not seek to reduce constraint violations. 

\subsection{Workload}
\subsubsection{Synthetic Workload} \label{synth-workload}
We used a synthetic workload (experiment sections D through F) where each job consists of meta-data, including job-id, resource requirement, and job duration. Jobs arrive online fashion controlled by $\lambda$, the job arrival rate. The higher the jobs arrival rate, the more jobs there are to process. We chose the job duration and resource requests such that $70\%$ of the jobs are short jobs and the remaining are long-duration jobs. Synthetic workload provides more nuanced control over simulation parameters allowing us to study the scheduler's behavior under a wide range of conditions~\cite{keynote-jsspp}~\cite{synth_workload}.

\subsubsection{HPC Workload} \label{workload}
We trained and evaluated the CoCoRL scheduler using Argonne National Laboratory (ANL) Intrepid HPC workload~\cite{ANL-intrepid} (experiment sections G and H). The logs contain several months' accounting records (from 2009) from the Blue Gene/P system called Intrepid. The ANL HPC workload is an old data set, but it has similar characteristics to modern workloads in terms of job arrival rates, resource requirements, and job duration. We made additional changes to the job logs to compensate for missing information. 

Jobs arrive in an online fashion, meaning that the scheduler does not know the job information \textit{a priori}. The jobs arrive at different rates; for instance, the $50\%$ job arrival rate means a new job arrives every other time step. The jobs vary in length, resource requirements, and value. The following experiments demonstrate the CoCoRL scheduler's performance with different job arrival rates. We used synthetic workload and power data for all the experiments in the following subsections. 

The real power prediction data (solar and wind) is from a existing Green\footnote{Reference omitted for double-blind review.} datacenter. The Green center is a microgrid equipped with 150 kW solar power and three wind turbines connected to the facility, each with 300 kVA of expected power generation. 

\subsection{Baseline Scheduling Policies}
The baseline heuristic scheduling policies, namely, Shortest-Job-First (SJF), First-Come-First-Serve (FCFS), QoS (Quality of Service), and Highest Value First (HVF), are compared with the CoCoRL scheduler in the experiments. The SJF, FCFS, and HVF are established and have intuitive definitions. The QOS scheduling policy schedules job with the highest QoS value. From \S\ref {jobs_qos}, the qos\_violation\_time specifies the upper bound by which the job must finish executing. The higher the QoS value of a job, the smaller qos\_violation\_time  (i.e., earlier deadlines). In essence, the QoS scheduling policy behaves like Earliest Deadline First (EDF) policy in that it schedules the jobs with lower qos\_violation\_time first. These greedy heuristic policies are chosen because each policy specifically optimizes a single metric (e.g., QOS minimizes delay, HVF maximizes job value).

\subsection{Performance Metrics}
The performance metrics used in this paper are Total Job Value, \S\ref{rewards}), Job Completion ratio, and System Utilization ratio. The job Completion ratio is the ratio of the jobs finished and the total jobs submitted during the simulation. System Utilization is the ratio of resources used and the total number of resources. 

The primary objective of the CoCoRL scheduler is to maximize the total job value, and the secondary objective is to minimize the overall job delays. The primary objective, total job value, is directly measured by calculating the value of the completed jobs. The secondary objective is measured indirectly through job completion and system utilization ratios. The rationale is that fewer jobs are delayed if more jobs are completed on time. Similarly, high system utilization may indicate more jobs running on the system, further reducing overall job delays.

\subsection{Performance - Total Job Value}
\begin{figure}[h]
\centering
\includegraphics[scale=0.60]{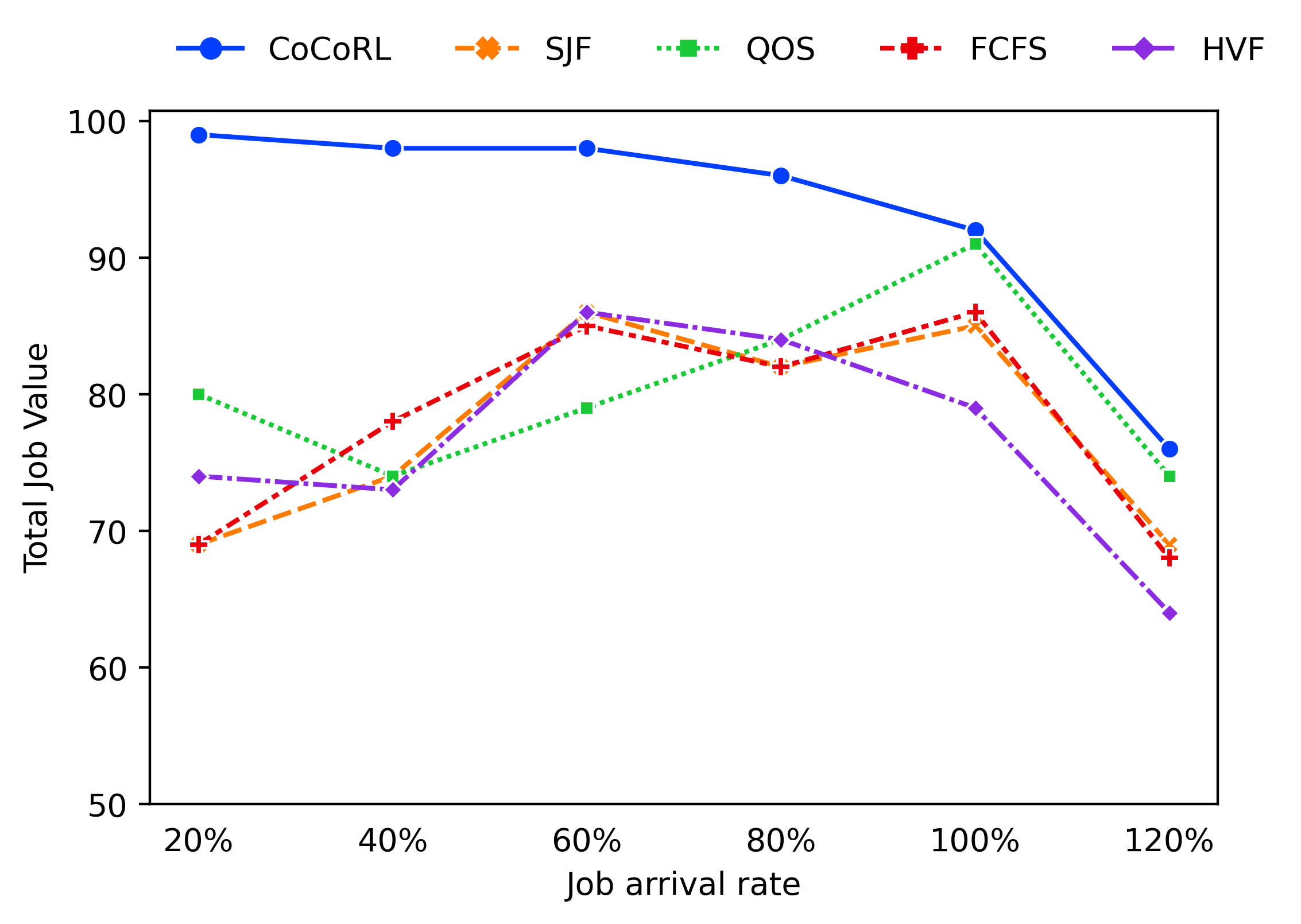}
\caption{Total job value with varying system load.} 
\label{fig:pid-job-val}
\end{figure}

This section evaluates the performance of different scheduling policies (CoCoRL, SJF, QOS, FCFS, and HVF) as the job arrival rate increases. Fig.\ref{fig:pid-job-val} shows performance in terms of the Total Job Value ratio, the total value of finished jobs, and the total value of all the jobs during the entire simulation and varying job arrival rates (between 20\% and 120\%). The 95\% confidence interval for each point of the Job Value ratio is within $\pm 0.025$. 

\textit{Analysis}: The primary objective of the CoCoRL scheduler is maximizing the total job value, which the CoCoRL scheduler accomplishes, illustrated in Fig.\ref{fig:pid-job-val}. The performance of the CoCoRL scheduler is significantly (20-28\%, 16-20\% and 12-18\%) better for arrival rates of 20\%, 40\%, and 60\%, respectively since there are fewer jobs (lower job arrival rate) than available resources in the datacenter. The total job value ratio gradually decreases as the job arrival rate increases (60-80\% arrival rate) due to more jobs than available resources. At 60-80\% job arrival rate, the CoCoRL performs 12-14\% better than baseline heuristic policies. As the system load increases (100-120\% arrival rate), CoCoRL and heuristic policies' performance drops because there is a significantly higher number of jobs than available resources, leading to a diminished job value ratio. The performance of CoCoRL is marginally better (1-6\%) than heuristic policies.

\subsection{Performance - Job Completion}
\begin{figure}[h]
\centering
\includegraphics[scale=0.60]{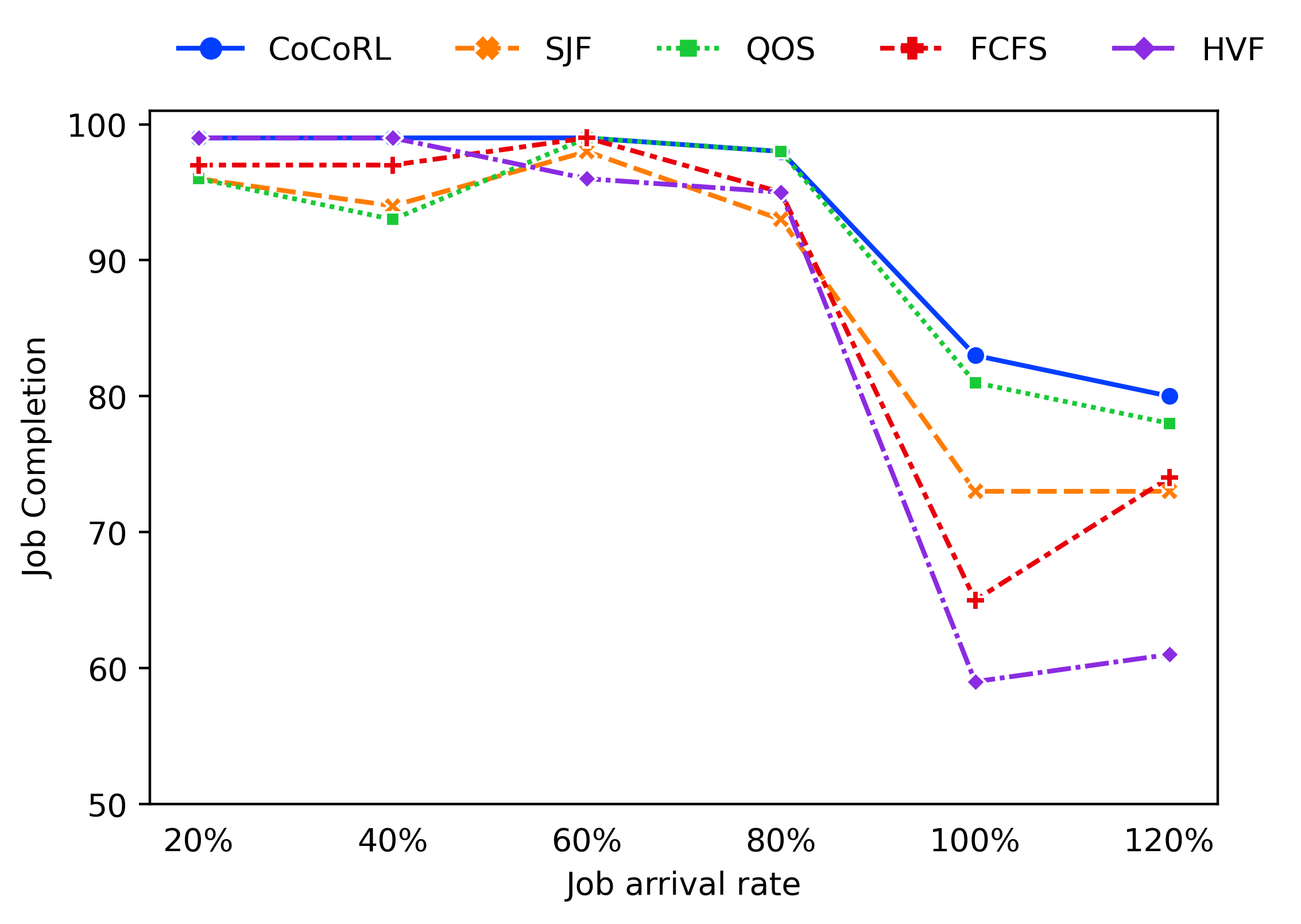}
\caption{Job completion ratio with varying system load.} 
\label{fig:pid-jobs-ratio}
\end{figure}

This section evaluates the performance of different scheduling policies (CoCoRL, SJF, QOS, FCFS, and HVF) and their job completion ratio as the job arrival rate increases (between 20\% and 120\%). Fig.\ref{fig:pid-jobs-ratio} shows performance in terms of job completion ratio, i.e., the number of jobs completed and the total number of jobs as the job arrival rate varies. The 95\% confidence interval for each point of the Job Completion ratio is within $\pm 0.015$.

\textit{Analysis}: From Fig.\ref{fig:pid-jobs-ratio}, the CoCoRL and heuristic policies demonstrate 93\% to 99\% job completion ratio when the system load is low (20-60\% arrival rate). As the system load increases (60-100\% arrival rate), the CoCoRL and QOS scheduling policies complete more jobs (about 97\%) than the other heuristic policies. When the system load is over 100\%, the job completion ratio drops for all the scheduling policies, including CoCoRL. We note that the CoCoRL scheduler performs comparably to the QOS policy but has around 5\% to 22\% more jobs than the other heuristic policies.

\subsection{Performance - System Utilization}

\begin{figure}[h]
\centering
\includegraphics[scale=0.60]{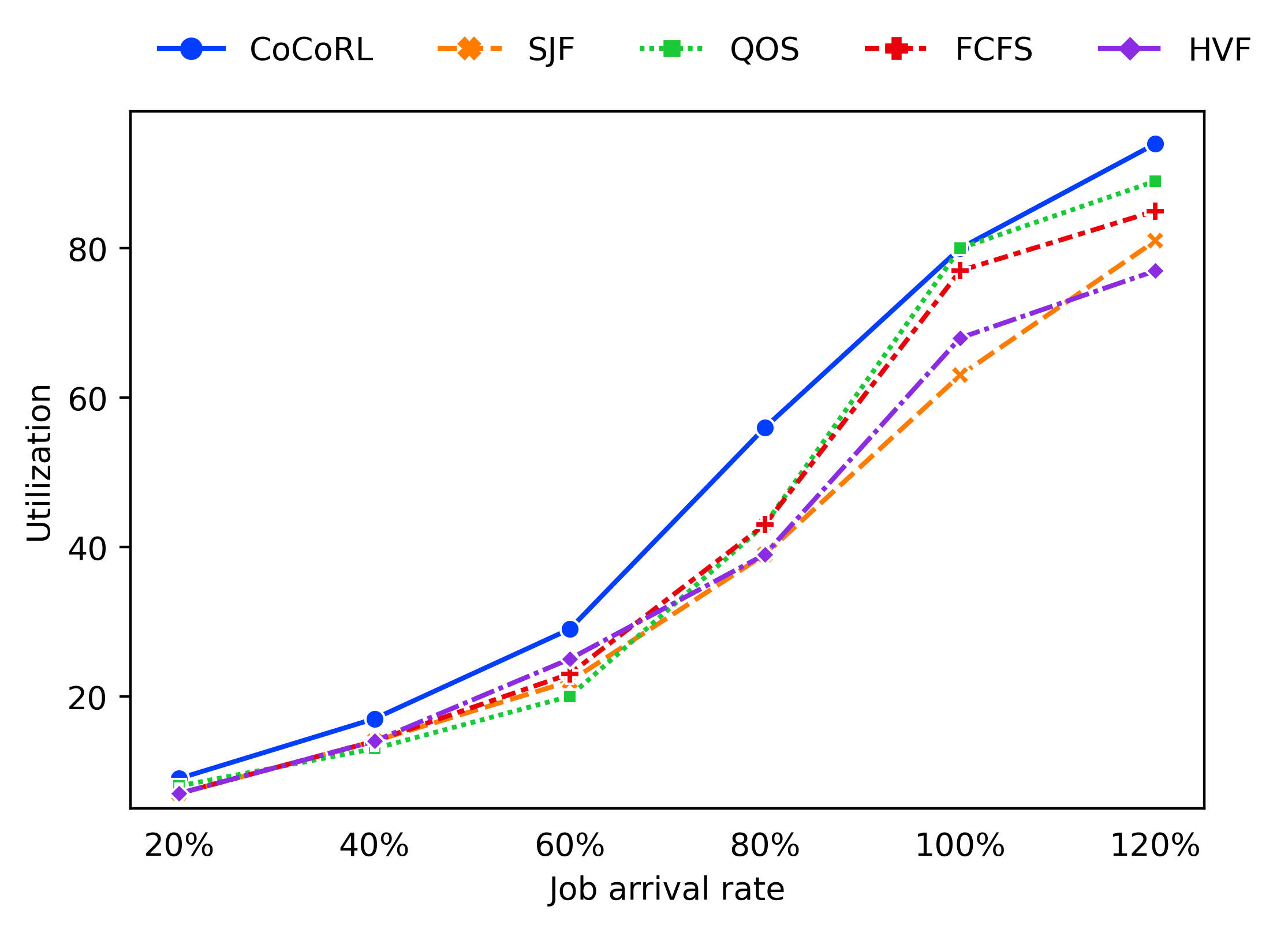}
\caption{System utilization ratio with varying system load.} 
\label{fig:pid-util}
\end{figure}

This section evaluates the system utilization using different scheduling policies (CoCoRL, SJF, QOS, FCFS, and HVF) as the job arrival rate increases. Fig.\ref{fig:pid-util} shows performance in terms of system utilization as the job arrival rate increases (between 20\% and 120\%). The 95\% confidence interval for each point of the Utilization ratio is within $\pm 0.01$.

\textit{Analysis}: From Fig.\ref{fig:pid-util}, the system utilization is low when the load on the system is low (20-60\% load). As the system load increases, the system utilization increases for all the scheduling policies, but CoCoRL provides a significantly higher utilization than heuristic policies. Since CoCoRL schedules more jobs (from the previous experiment), more resources are used, leading to higher system utilization. Even though more jobs are available as the system load increases (60\% or higher), it is possible that some of the jobs' resource requirements cannot be satisfied, fragmenting the resources. The fragmented resources lead to the underutilization of the system. Therefore, we do not see 100\% utilization even when the system load is 120\%. 

\begin{figure}[h]
\centering
\includegraphics[width=\linewidth]{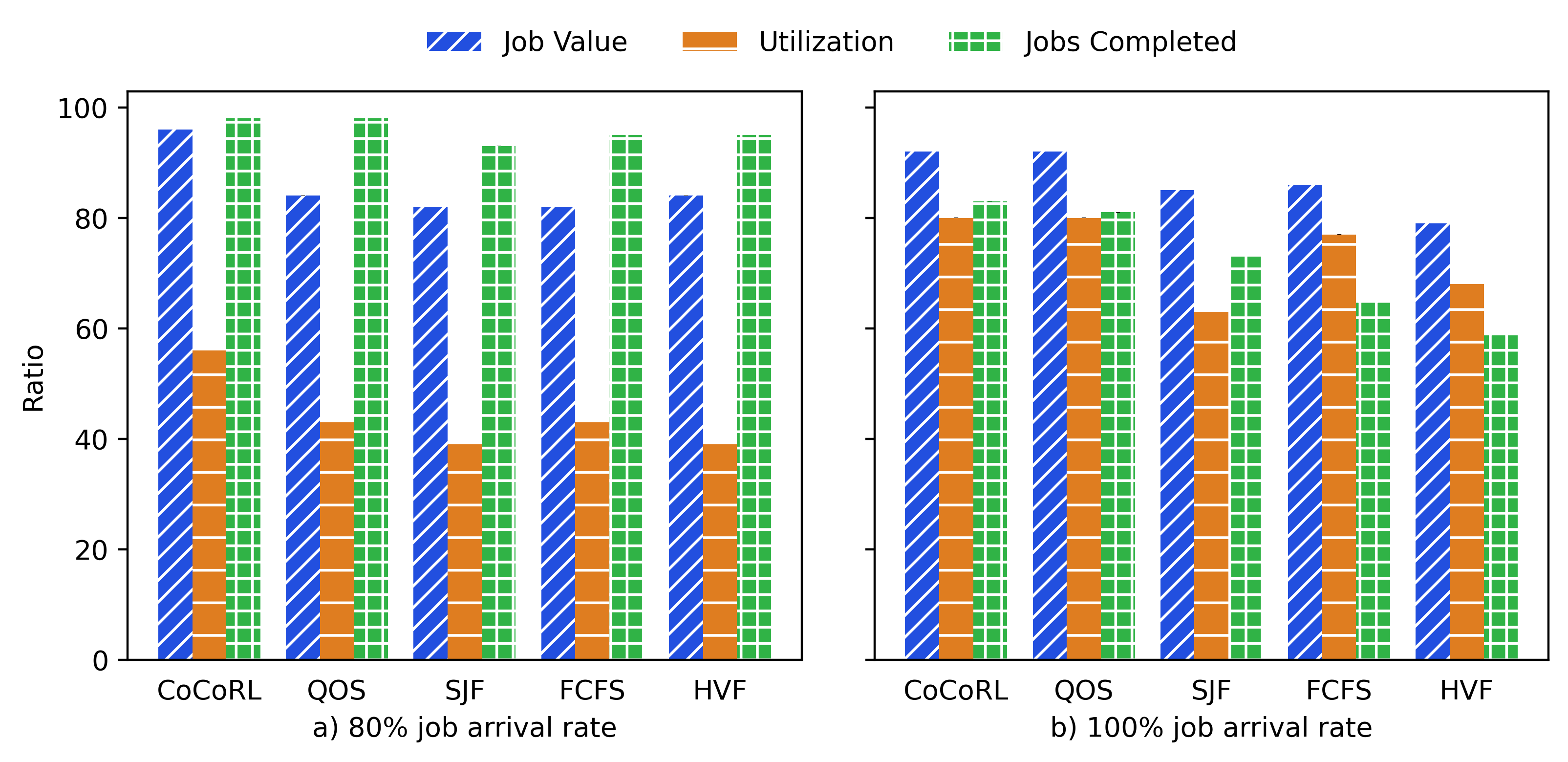}
\caption{Job Value, Jobs Completed and System Utilization at 80\% job arrival rate.} 
\label{fig:pid-comps}
\end{figure}

Finally, we compare the total job value, completion, and system utilization ratio for different scheduling policies (CoCoRL, SJF, QOS, FCFS, and HVF) at $80\%$ and $100\%$ job arrival rates, plotted in Fig.\ref{fig:pid-comps}. 

In summary, Fig.\ref{fig:pid-comps} shows that CoCoRL can simultaneously achieve (1) a higher total job value, (2) high system utilization, and (3) a high job completion ratio. We show CoCoRL performing considerably better in all of the above categories at 80\% job arrival rate (Fig.\ref{fig:pid-comps}.a). At 100\% job arrival rate (Fig.\ref{fig:pid-comps}.b), CoCoRL performs at par or better than the QOS heuristic scheduling policy. None of the other heuristic policies under comparison meet these goals simultaneously. For a job arrival rate above 100\%, we note that the QOS scheduling policy is the second best policy (after CoCoRL) for simultaneously satisfying all of the above goals. 

\subsection{Performance with ANL HPC workload}
This section evaluates the performance of different scheduling policies (CoCoRL, SJF, QOS, FCFS, and HVF) as the job arrival rate increases with the ANL HPC workload. 

Fig.\ref{fig:pid-job-val} shows performance in terms of the Total Job Value ratio, the total value of finished jobs, and the total value of all the jobs during the entire simulation and varying job arrival rates (between 20\% and 120\%). The 95\% confidence interval for each point of the Job Value ratio is within $\pm 0.02$. 

\begin{figure}[h]
\centering
\includegraphics[scale=0.60]{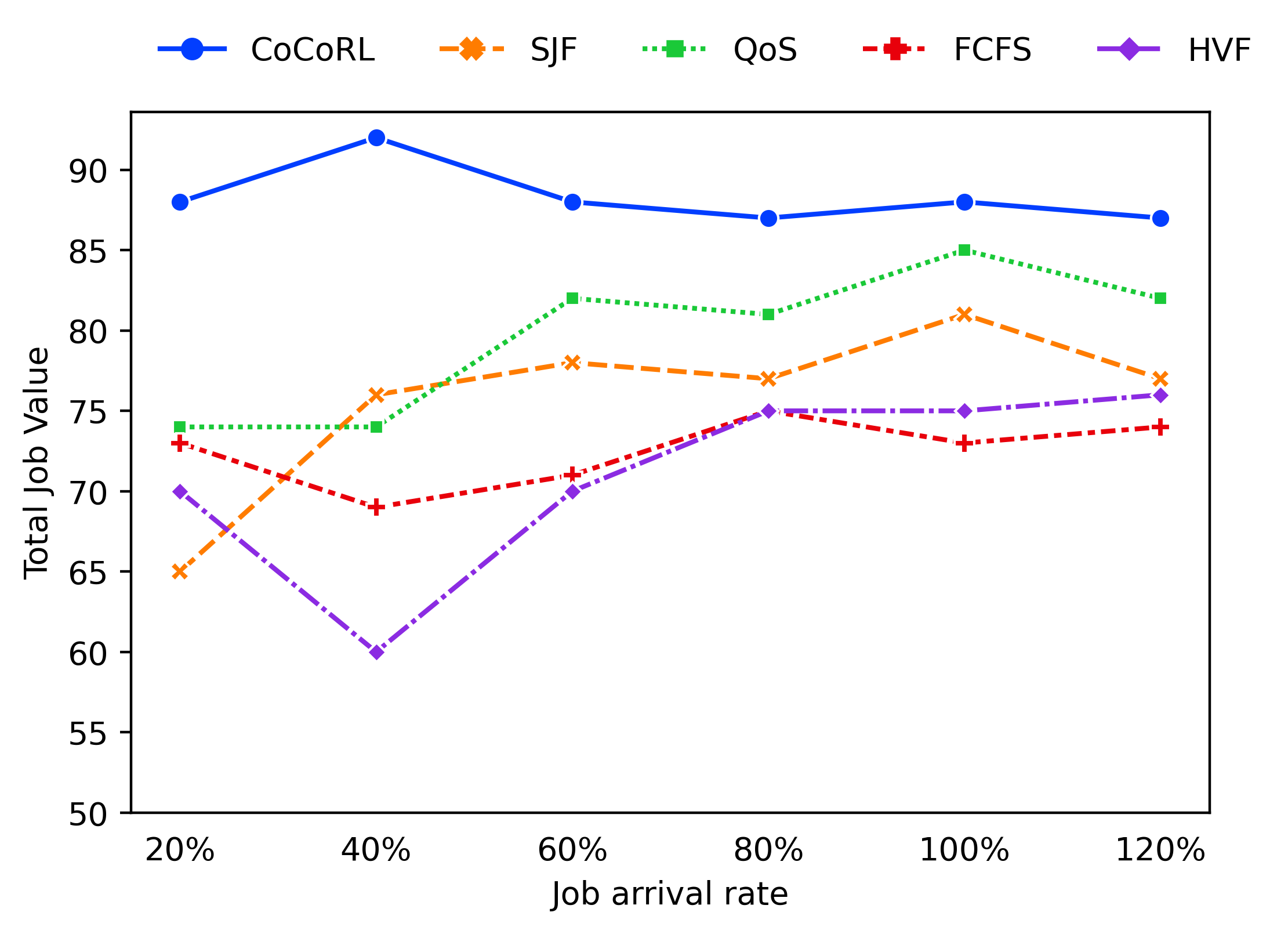}
\caption{Total job value with varying system load - ANL HPC workload.} 
\label{fig:real-value}
\end{figure}

\textit{Analysis}: Fig.\ref{fig:real-value} shows the performance of the CoCoRL which is significantly (11-23\%, 16-22\%, and 10-18\%) better for arrival rates of 20\%, 40\% and 60\% respectively. At 80\% job arrival rate, the CoCoRL performs 6-12\% better than baseline heuristic policies. As the system load increases (100-120\% arrival rate), CoCoRL's performance maintains a moderately higher (5-10\%)  total job value than heuristic policies. All scheduling policies perform consistently similarly as the job arrival rate increases, the artifact of ANL job characteristics as most of the jobs in this workload are big jobs (higher resource requirement and longer job duration).

\begin{figure}[h]
\centering
\includegraphics[scale=0.55]{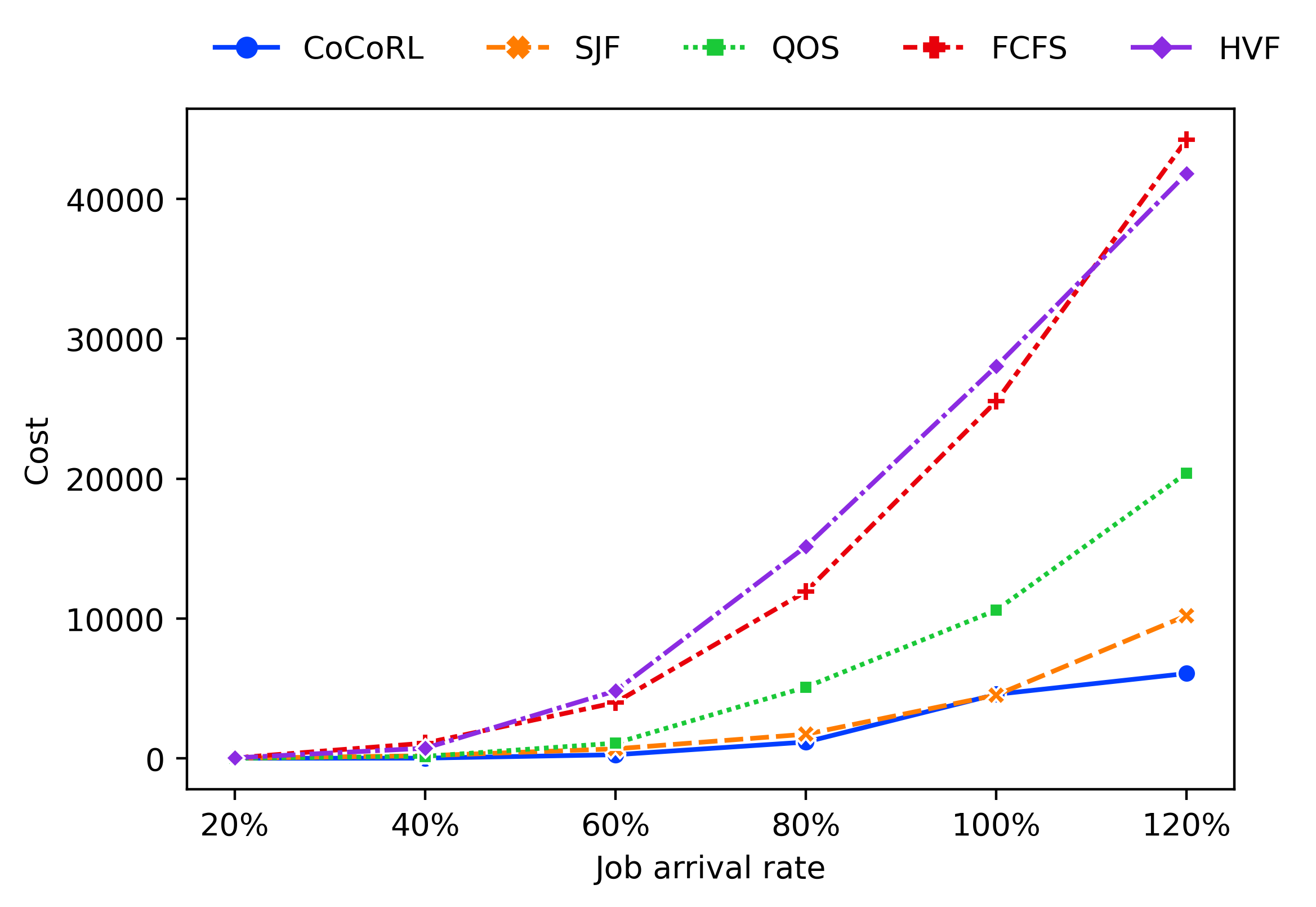}
\caption{Accrued cost with varying job arrival rates - ANL HPC workload.} 
\label{fig:real-cost}
\end{figure}

From ~\S\ref{rewards}, the environment provides a separate cost signal for each delayed job at every timestep. The cost is separate from the reward signal. Fig.\ref{fig:real-cost} shows the performance of the CoCoRL in terms of total cost accrued during the entire simulation with varying job arrival rates (between 20\% and 120\%). The 95\% confidence interval for each point of the Job Value ratio is within $\pm 0.25$. For this experiment, the cost\_limit was set at 2000.

\textit{Analysis}: Fig.\ref{fig:real-cost}, at 20\% job arrival rate, CoCoRL, SJF and QOS all accrue similar costs whereas FCFS and HVF accrue higher costs. We note that CoCoRL accrues significantly lower costs for job arrival rates 0f 40-80\%. CoCoRL and SJF maintain a meager accrued cost as the job arrival rate increases, while the other heuristic policies accrue significantly higher costs. 

\subsection{Hyperparamaters - PID and Cost Limits}
In this subsection, we will highlight the significance of tunning parameters, namely values for $K_p,\ K_i,\ and\ K_d$ shown in the training algorithm.\ref{algo:cppo-training}. We note that the values of $k_p, K_i$, and $K_d$ are tuned so that they satisfy the datacenter optimization requirements. 

\begin{figure}[h]
\centering
\includegraphics[scale=0.6]{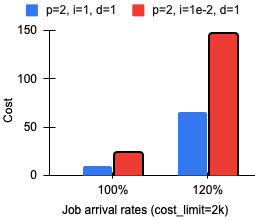}
\caption{Cost with different PID values - ANL HPC workload.} 
\label{fig:pid-cost}
\end{figure}

\textit{Analysis}: Fig.\ref{fig:pid-cost} shows the performance of the two models trained with different PID values ($K_p=2, K_i=1, K_d=1$ and $K_p=2, K_i=1e-2, K_d=1$). Model with higher integral value, $K_i=1$ maintain significantly lower costs compared to lower integral value, $K_i=1e-2$ for job arrival rates of 100-120\%. The high $K_i$ setting generally achieves better cost performance due to efficient cost control.

In addition to PID values, setting the appropriate cost\_limit ensures that the scheduler produces the desired result within specified limits. Cost limit was identified by running the unconstrained RL and accumulating cost signal for delayed jobs. We reduced the cost limit from the previous step so that CoCoRL can make reasonable trade-off between rewards and costs.

\begin{figure}[h]
\centering
\includegraphics[scale=0.6]{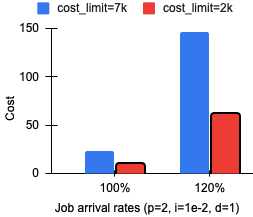}
\caption{Cost with different cost\_limit - ANL HPC workload.} 
\label{fig:cl-cost}
\end{figure}

\textit{Analysis}: Fig.\ref{fig:pid-cost} shows the performance of the two models trained with different cost\_limit values (7000 and 2000). For example, the datacenter operator can set conservative cost\_limit to ensure QoS violations are minimized. On the other hand, selecting a lenient cost\_limit might generate a higher total job value where the scheduler picks higher-value jobs at the risk of violating the QoS agreements of low-value jobs. In essence, if the cost limit is too small, the scheduler agent will learn unsafe scheduling behaviour leading to QoS violations, and if the cost is too severe, the agent may fail to learn anything useful.

%% file: 7-related-work.tex
\section{Related Work} \label{related_work}

\subsection{Scheduling using heuristics and feedback control}
This system in ~\cite{old-gdc-2013} employs a heuristic Mixed-Integer Linear Programming (MILP) formulation to minimize grid electricity when the electric grid is used and minimize the workload's performance degradation when using renewable energy sources. A wide variety of contributions to the design and implementation of resource management in various settings using control theory ~\cite{feedback_control_qos}~\cite{feedback_control_iot}~\cite{feedback_control_mete}. The scheduler in~\cite{stankovic-99} applies the PID controller to load balance tasks in real-time systems. All of these systems use heuristics-based scheduling algorithms in their design. Hand-engineering domain-specific heuristics-based schedulers to meet specific objective functions is time-consuming, expensive, and needs extensive tuning in dynamic environments.

\subsection{RL schedulers with single objective}
Much of the  prior work focuses on unconstrained optimization problems with no safety constraint, i.e., optimizing a single objective such as latency, throughput, power, and energy consumption. The Spotlight~\cite{google-spotlight} partitions the agent's neural network training operations onto different devices (CPUs and GPUs) to minimize model training time. The RL scheduler in~\cite{Spear-ICDCS19} is designed to minimize the makespan of DAG jobs considering both task dependencies and heterogeneous resource demands. The scheduler in~\cite{RLScheduler-JSSPP2021} implements a co-scheduling algorithm based on an adaptive RL by combining application profiling and cluster monitoring. The optimization objective in~\cite{RLScheduler-JSSPP2021} is to maximize resource utilization. These implementations incorporate a single reward function that maximizes or minimizes a single objective. Although effective, this naive application of RL to achieve a single objective can lead to poor performance on other objectives. 

\subsection{Multi-objective schedulers}
Numerous prior work has studied multi-objective optimization in scheduling~\cite{survey_buyya}. Recent work~\cite{datacenter-energy-cdc} presents a unified management approach for the thermal and workload distribution in data centers, modeled as a Model Predictive Control (MCP) problem. The objective is to minimize power consumption while satisfying thermal and Quality of Service (QoS) constraints. They implement a particle-based algorithm that requires adjusting some non-trivial number of parameters over multiple iterations. DeepEE~\cite{DeepEE} proposes improving datacenters' energy efficiency by concurrently considering the job scheduling and cooling systems. The goal, in~\cite{DeepEE}, is to reduce cooling costs in a datacenter rather than optimize job scheduling. The joint optimization problem aims to minimize the power usage effectiveness while preventing overheating in the rack server and keeping the load balance. The downside of these systems is that designing a good reward function that balances different, often conflicting, objectives is challenging. For example, a different optimal solution exists for each set of penalty coefficients, also known as Pareto optimality. Manually tuning the exact coefficients of different objectives is a time-consuming process.

\subsection{RL with constraints}
A good number of prior work has studied safety constraints in RL training~\cite{saferl_survey}. Two main categories of the work include limiting the action space or state space to satisfy constraints. The work in~\cite{saferl_shielding} blocks specific actions when a safety condition is violated. However, shielded RL only applies to simple problems with tabular states and treats the shield as part of the environment without modifying the RL optimization objective. Therefore, the shield is only a protection mechanism to constrain the action set, not a technique for adapting the agent's policy. The proposed work in~\cite{saferl_pid} uses the RL framework for robotics to inject experience data from the expert's control into the replay buffer for off-policy RL methods. This framework uses a classical controller (e.g., a PID controller) as an alternative to speed up the training of RL for robot planning and navigation problems. As the actor-network becomes more advanced, it can then take over to perform more complex actions. Eventually, the PID controller is discarded entirely. 

Another body of prior work focuses on adjusting RL algorithms to avoid exploring unsafe states~\cite{saferl_bootstrap}, modeling the risk in state transitions, and conservative exploration~\cite{saferl_state_control}. These methods incorporate the fallback policy within neural networks, but they cannot guarantee that the system recovers to a safe state eventually. Training wheels~\cite{safe_load_balancer} prioritize meeting the safety condition and rely on deterministic fallback policies. While it may not be possible to impose limits directly by prescribing constraints in the action or state space; instead, hazard-avoiding behavior must be learned. Our work focuses on embedding all conflicting requirements in a constrained RL problem and learning the parameters automatically. Our datacenter simulator is akin to the model presented in~\cite{RLScheduler-JSSPP2022}. 

%% file: 6-discussion.tex
\section{Future work} \label{discussion}

We hand-tuned hyperparameters to attain our optimization goals. However, we note that our hand-tuning is not indicative of the best possible performance of each of our optimization goals. We plan to explore the tuning hyperparameters for other multi-criteria optimization problems. For example, selecting from various power sources to minimize brown energy and battery usage. Multi-criteria optimization is ongoing, and we will cover this topic in future work.

Additionally, we plan to investigate CoCoRL's performance under dynamic conditions, including system load and intermittent power supply in the green datacenter environment.

%% file: 8-conclusion.tex

\section{Conclusion} \label{conclusion}

This paper presented a Constraint-controlled RL (CoCoRL) scheduler that automatically learns conflicting reward and cost functions. We applied proportional-integral-derivative Lagrangian methods in Deep Reinforcement Learning to job scheduling problem in the green datacenter environment. We showed that a Constraint-controlled RL scheduler learns policies that satisfy the conflicting constraints and optimize for multiple objectives such as resource utilization and job completion. Our experiments demonstrate the CoCoRL scheduler's performance for both the primary objective (maximizing total job value) and the secondary objective (minimizing overall job delay). The CoCoRL scheduler yields significantly higher total job value while exhibiting comparable job completion and system utilization ratio than baseline heuristic scheduling policies. We showed that our CoCoRL scheduler is robust and efficiently adapts to HPC workloads and power supply data (solar and wind) from an existing Green datacenter. Additionally, we showed the significance of accurately tuning hyperparameters to satisfy various optimization goals set by datacenter operators.